\documentclass[a4paper,11pt,nofootinbib]{revtex4-1}
\usepackage{listings, color}  
\usepackage{amsmath, bm}  
\usepackage{graphicx}
\usepackage{tipa}
\usepackage{url}
\usepackage{comment}

\begin{document}
\title{Individual External Dose Monitoring of All Citizens of Date City by Passive Dosimeter 5 to 51 Months After the Fukushima NPP Accident (series):\\
1. Comparison of Individual Dose with Ambient Dose Rate Monitored by Aircraft Surveys}
\author{Makoto Miyazaki}
\affiliation{Department of Radiation Health Management, Fukushima Medical University, Fukushima 960-1295, Japan
}
\author{Ryugo Hayano}
\affiliation{Department of Physics, The University of Tokyo, Tokyo 113-0033, Japan
}

\begin{abstract}
Date (d\textschwa 'te) City in Fukushima Prefecture has conducted a population-wide individual dose monitoring program after the Fukushima Daiichi Nuclear Power Plant Accident, which provides a unique and comprehensive data set of the individual doses of citizens. 
The purpose of this paper, the first in the series, is to establish a method for estimating effective doses based on the available ambient dose rate survey data.
We thus examined the relationship between the individual external doses and the corresponding ambient doses assessed from airborne surveys. The results show that the individual doses were about 0.15 times the ambient doses,  the coefficient of 0.15 being a factor of 4 smaller than 
 the value employed by the Japanese government, throughout the period of the airborne surveys used. 
 The method obtained in this study could aid in
the prediction of individual doses in the early phase of future radiological accidents involving large-scale contamination.
\end{abstract}
\keywords{Fukushima Daiichi accident, radiation protection, dose assessment, individual dose,  airborne survey, ambient dose, Date City.}
\maketitle

\section{Introduction}

In making decisions regarding appropriate protection policy against radioactive contamination after nuclear accidents (such as the Fukushima Daiichi Nuclear Power Plant: FDNPP accident), obtaining the individual  dose distribution is crucial. The ICRP recommends~\cite{icrp103,icrp111} the responsible bodies to characterize the individual dose distribution of the exposed population, and to take appropriate and optimized actions based on the ALARA principle.

The external doses in the first four months of the Great East Japan Earthquake were estimated within the framework of  the basic health survey conducted by the Fukushima Prefectural government, based on population movement and activities extracted from the questionnaire distributed to Fukushima residents~\cite{ishikawa}. 

After the summer of 2011, many Fukushima municipalities started individual external dose monitoring for members of the public living in existing exposure situations, although such large-scale measurements of the general public have not been considered mandatory in the conventional radiation protection schemes. 
These municipal dose measurement programs were not initiated nor supervised by the Japanese central government, and thus they were not standardized in terms of the target group, 
distribution/collection methods, 
the format of  data dissemination,
and
instructions describing how the dosimeters should be used. The information available to both the residents and administration in the affected areas is therefore rather fragmented. This has made it difficult for citizens to grasp the overall picture of the external exposures even though data have been published by each municipality (e.g., Fukushima City~\cite{fukushimagb}) and by academic sectors~\cite{tsubokura2015}. As such, the collected data have not been well utilized for the benefit of the affected communities.

Date  City is located about 50--60\,km northwest of FDNPP. Unlike Iitate Village, which is adjacent to Date City and from which the entire population was ordered to evacuate in April 2011 (called the ``deliberate evacuation area'' \cite{evacuation}), no evacuation order was issued to Date residents, except for a limited number  (128) of households  for which the Japanese Government estimated that the exposure to the residents would exceed 20\,mSv within the first year after the accident (called the ``specific spots recommended for evacuation'' \cite{evacuation}). 
Among the municipalities in which all the residents continue to live after the FDNPP accident, Date City has the largest gradient of contamination levels within the city borders. 

In Date City, residents voluntarily started and actively participated in radiation protection measures immediately after the accident, e.g., ambient dose measurements and school decontamination. The city office also endorsed and supported those activities from early on~\cite{datecityweb}. The Mayor of Date City ordered the launch of a project to monitor the individual doses of residents, as well as the ambient dose rate throughout the city, with an emphasis on gathering information which could be used to prioritize decontamination based on the contamination levels. 

For the measurement of individual external doses,  Date City distributed individual dosimeters (radio-photoluminescence (RPL) glass dosimeters: Glass Badge\textsuperscript\textregistered) to kindergarten-, elementary- and junior high school-children in August 2011. The target group was subsequently enlarged as the production capacity of the supplier increased,  and the measurements are still ongoing. 

It is very notable that Date City distributed the glass badges to all citizens ($\sim 65,000$ in all) for one year, from July 2012 to June 2013~\cite{datefukko}, foreseeing that the actual dose data would become required when communicating to the citizens. The results published by Date City were further analyzed by the IAEA~\cite{iaea}.
No other external dose measurement program of comparable scale has been done in Fukushima, making Date City's data the most comprehensive set of actual measurements of individual  doses carried out after the FDNPP accident.

In addition to the external doses, Date City  also started to measure the internal doses 
of citizens in October 2011, which still continue. The decontamination of public facilities and roads was started in the fall of 2011, and that of residences was started in the fall of 2012 (to be discussed further in the following section). In March 2015, Date City announced the completion of its decontamination program.

The present authors made use of the large-scale individual dose monitoring data provided by Date City covering the period from 5 to 51 months after the FDNPP accident,  analyzed the relationship of the individual doses to the results of airborne surveys conducted by the Japanese Government~\cite{nramonitoring}, the effect of decontamination on the individual doses, and the relationship between the external and internal doses. These results will be published in a series of three papers.

The purpose of this paper, the first in the series, is to establish a method for estimating individual external doses based on the available ambient dose rate survey data, such as the airborne surveys.

\section*{ETHICS STATEMENT}
Date city  mayor's office entrusted the data to the authors, one of whom (MM) is a Date municipal government advisor. The geocoded household addresses of the glass-badge monitoring participants were pseudo-anonymized by rounding both longitude and latitude coordinates to 1/100 degrees prior to data analyses.
This study was approved by the ethics committee of the Fukushima Medical University (approval No. 2603).

\section{MATERIALS AND METHODS}
\subsection{The terrain and radioactive contamination situation of Date City
}

Date City is located about 50--60\,km northwest of FDNPP. Adjacent to Date City, the prefectural capital Fukushima City lies to the southwest, Kawamata Town lies to the south, and Iitate Village lies to the southeast. The total area is about 265 square kilometers, of which agricultural land comprises about 27\%, forest about 38\%. Date City is  endowed with a rich natural environment and fertile arable land. The population in 2010 (prior to the disaster) was about 67,000, and  the number of households was about 20,000, most of which were unevenly distributed in the flat land towards the northwest of  the city.
The population density is low in the eastern and southern parts, which is  mountainous, being a part of the Abukuma-kochi highland. 
About 5,400 households (26\%) engage in agriculture, of which 900 are full time farmers. Their revenues come mostly from fruit production~\cite{datestat}. This, together with the large fraction of agricultural land area, means that agriculture is an essential part of life in Date City. 

\begin{figure}
\begin{center}\includegraphics[width=\textwidth]{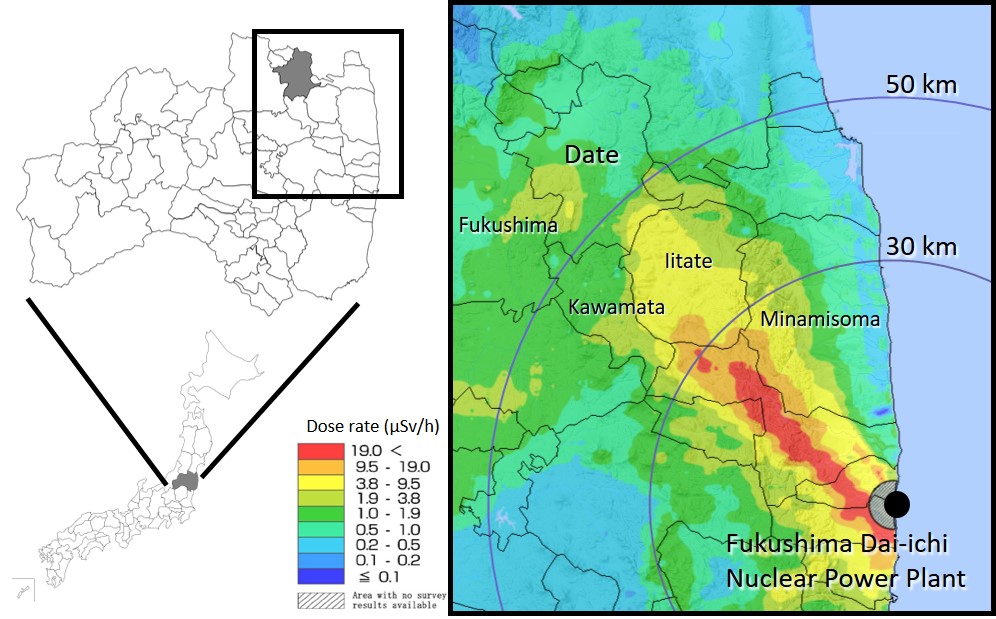}\end{center}
\caption{\label{fig1} Maps showing the location of Date City and the distribution of ambient dose rates as of November 5, 2011, based on the airborne monitoring surveys conducted from October to November 2011. Adopted from a map published in Ref.~\cite{nraairborne}.}
\end{figure}

Radioactive materials released by the FDNPP accident were unevenly deposited in the northwesterly direction from FDNPP as shown in Figure \ref{fig1}. As a result, the contamination levels have a strong gradient from south (high) to north (low) in Date City.

\subsection{Zoning based on the difference in the contamination level in Date City}

In order to assess the distribution of the ambient dose rate throughout the city, in August 2011, Date City generated a map of the ambient dose rate measured at 1m above the ground by dividing the city into a 1\,km grid (500\,m in urban areas). Based on this map, Date City made a decontamination plan (first edition) in October 2011, which stated that decontamination method would be  optimized according to the contamination level~\cite{datedecontam1}. 

\begin{figure}
\begin{center}\includegraphics[width=0.7\textwidth]{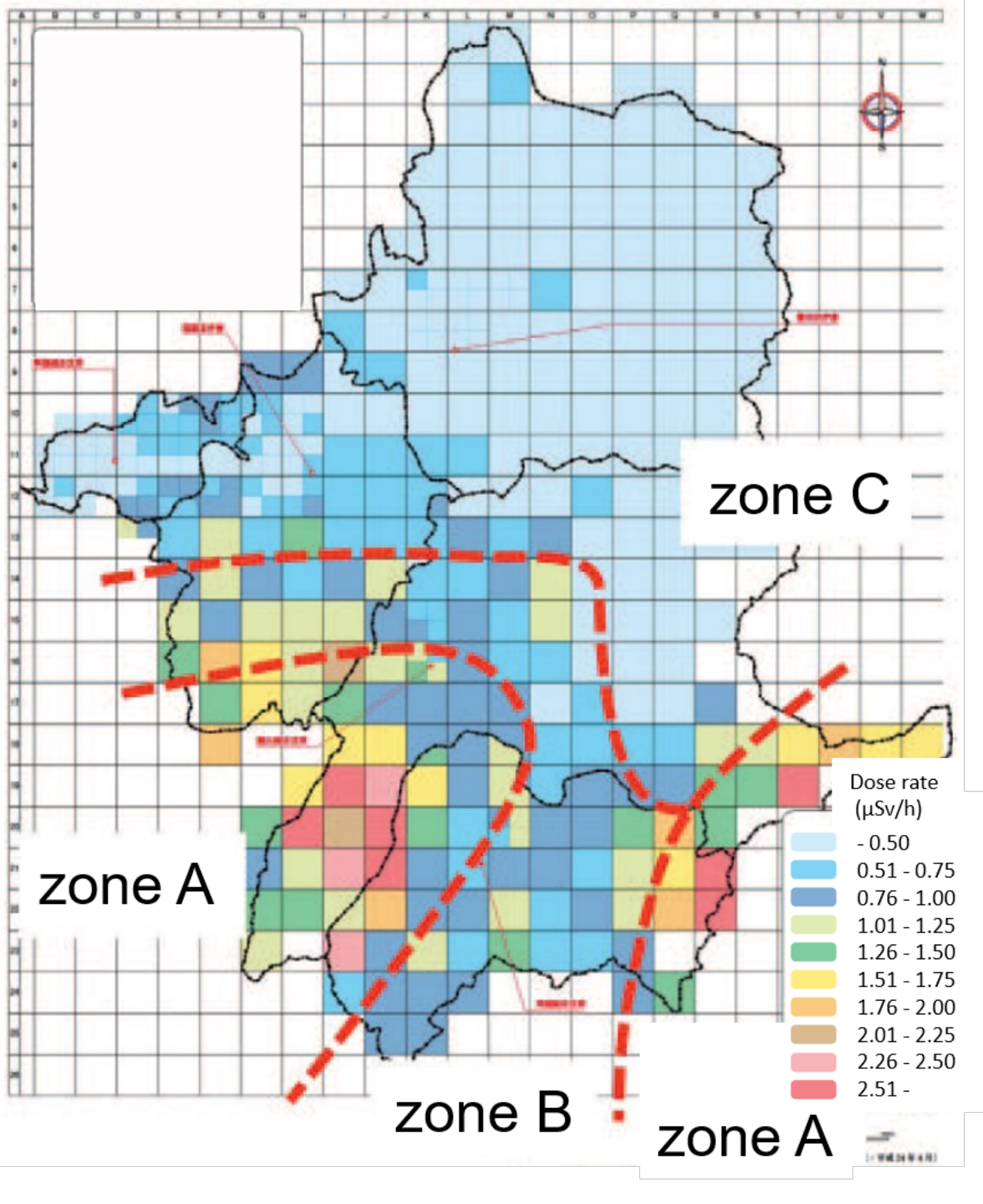}\end{center}
\caption{\label{fig2} 
The ambient dose rate map of Date City measured in March 2012, together with the zoning according to the dose rate (as measured in August 2011): zone A ($>3.5\,\mu$Sv\,h$^{-1}$), zone B ($1\, \mu$Sv\,h$^{-1}$ $\sim 3.5\, \mu$Sv\,h$^{-1}$) and zone C ($< 1\,\mu$Sv\,h$^{-1}$).  Adopted from a map published in Ref.~\cite{datecityweb}.
}
\end{figure}

In the Date City decontamination implementation plan (2nd edition) published in August 2012, a zoning scheme was introduced~\cite{datedecontam2}. Date City used the ambient dose map of August 2011 and the formula used by the Japanese government (see Ref.~\cite{safetyassessment}, and Eq.~\ref{eq:jgf}), to classify the households where the annual external dose was estimated to exceed 20 mSv y$^{-1}$ as A, less than 5 mSv y$^{-1}$ as C, and the rest as B. 
In terms of the ambient dose rate, these corresponded to:  zone A ($\gtrsim 3.5\,\mu$Sv\,h$^{-1}$), zone B ($1\, \mu$Sv\,h$^{-1}$ $\sim 3.5\, \mu$Sv\,h$^{-1}$) and zone C ($< 1\,\mu$Sv\,h$^{-1}$). The three zones in Date City are shown in Fig. \ref{fig2}\footnote{Note that this map, made available to the authors by Date City is for the measurement carried out in March 2012, so that the dose rates are lower than in August 2011.}. The decontamination was to be carried out in this prioritized order. The numbers of households in each zone in August 2012 were as follows: 
\begin{eqnarray*}
\mbox{zone A} &\sim& 2,500 \mbox{~households},\\
\mbox{zone B} &\sim&  3,700 \mbox{~households, }\\
\mbox{zone C} &\sim& 15,600 \mbox{~households}.
\end{eqnarray*}

\subsection{Individual dose monitoring of residents using  glass badges}

The individual dose monitoring of residents in Date City was carried out using glass badges usually used for radiation workers; their use by residents living in contaminated areas was not foreseen before the FDNPP accident.

Glass badges are calibrated  against personal dose equivalent Hp(10) on a polymethyl methacrylate (PMMA) slab phantom in the anterior-posterior (AP) geometry~\cite{icru74}.
However, residents living in contaminated areas receive radiation close to the rotational irradiation geometry, as the radioactive cesium released in the FDNPP accident now exists almost uniformly in the environment.
Even in this case, it has been shown that a dosimeter worn on the body trunk gives good approximation of the effective dose~\cite{hirayama2013}. 

The glass badge is sensitive to natural background radiation (terrestrial $\gamma$-rays and cosmic rays). In order to evaluate the additional dose due to the radioactive cesium released by the accident, it is necessary to subtract the background contribution. In generating the glass badge readout report for Fukushima Prefecture residents, the glass-badge supplier, Chiyoda Technol, subtracts a value equivalent to 0.54\,mSv y$^{-1}$ as the background, measured at Oarai Town, Ibaraki Prefecture (130\,km south of FDNPP), where the contamination due to the FDNPP accident was low~\cite{nomura2015}. Date City then mails the results to each monitoring survey participant.

\subsection{Monitoring survey participants and measuring periods}

The first external contamination measurement survey in Date was conducted for one month in August 2011, and targeted pregnant women and children aged 15 or under. 
In all subsequent measurement surveys, the measurement period was three months (one quarter, hereafter denoted Q). In particular, the measurement survey conducted from Q2 2012 (Japanese fiscal year, i.e, July-September 2012) through Q1 2013 targeted  all citizens, $\sim 65,000$ in all, of which 81\% (52,783) received and returned the badge every 3 months for one year. After that, all children up to the age of 18, all pregnant women, and all persons living in zone A continued to participate in the measurements, while the number of participants living in zone B and C has been gradually reduced; in these zones, glass badges are distributed based on random sampling of the population, and also to residents who request one, instead of targeting the entire population.

Table 1 shows the timeline of the glass-badge measurements through Q1 2015, the change of subjects,  age distribution, and the total number of participants upon which the present analysis is based.

\begin{table}
\caption{\label{tab1} Timeline of the glass-badge measurement surveys in Date City from 5 to 51 months after the FDNPP accident.}
\begin{center}\includegraphics[width=\textwidth]{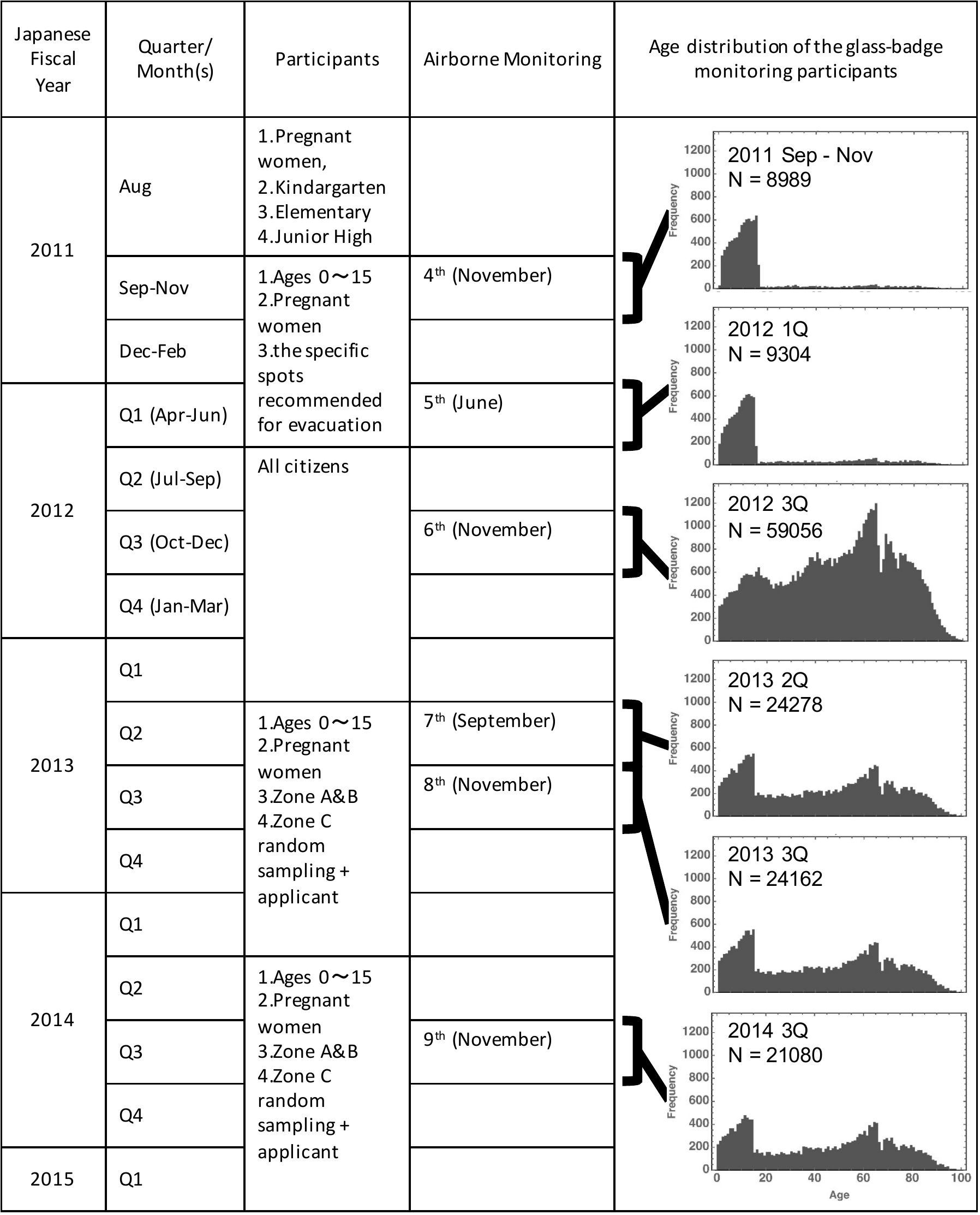}\end{center}
\end{table}

\subsection{Method for comparing   individual doses and  ambient doses}

\begin{figure}
\begin{center}
\begin{minipage}{0.44\textwidth}
\includegraphics[width=.78\textwidth]{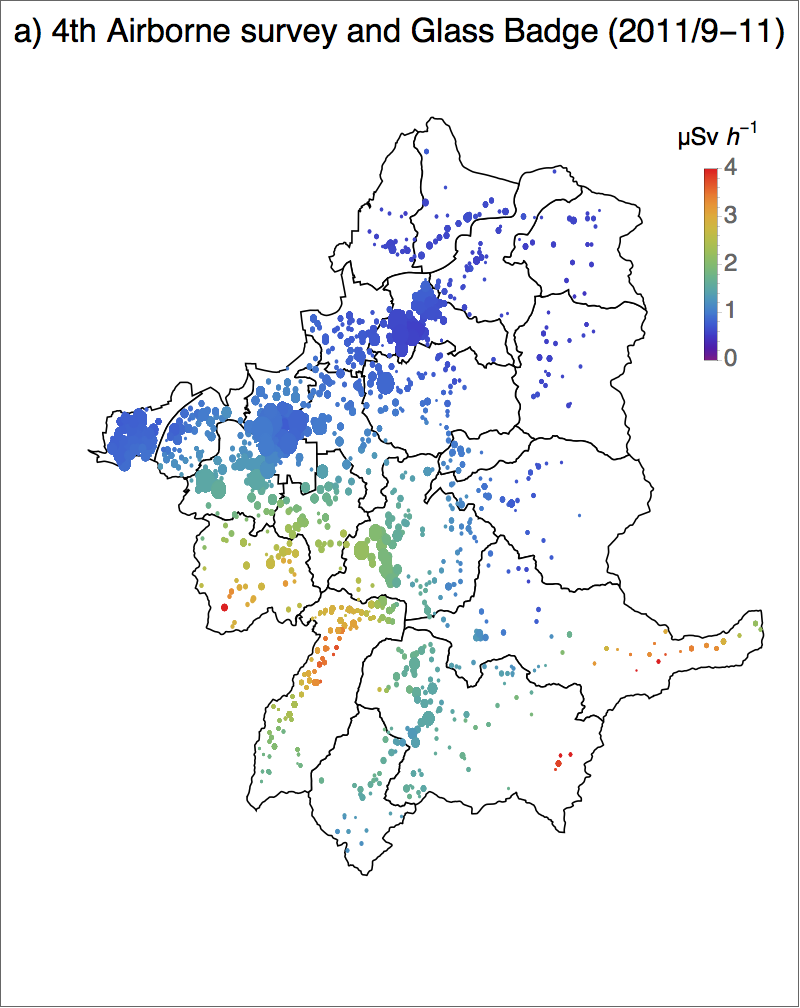}
\includegraphics[width=.78\textwidth]{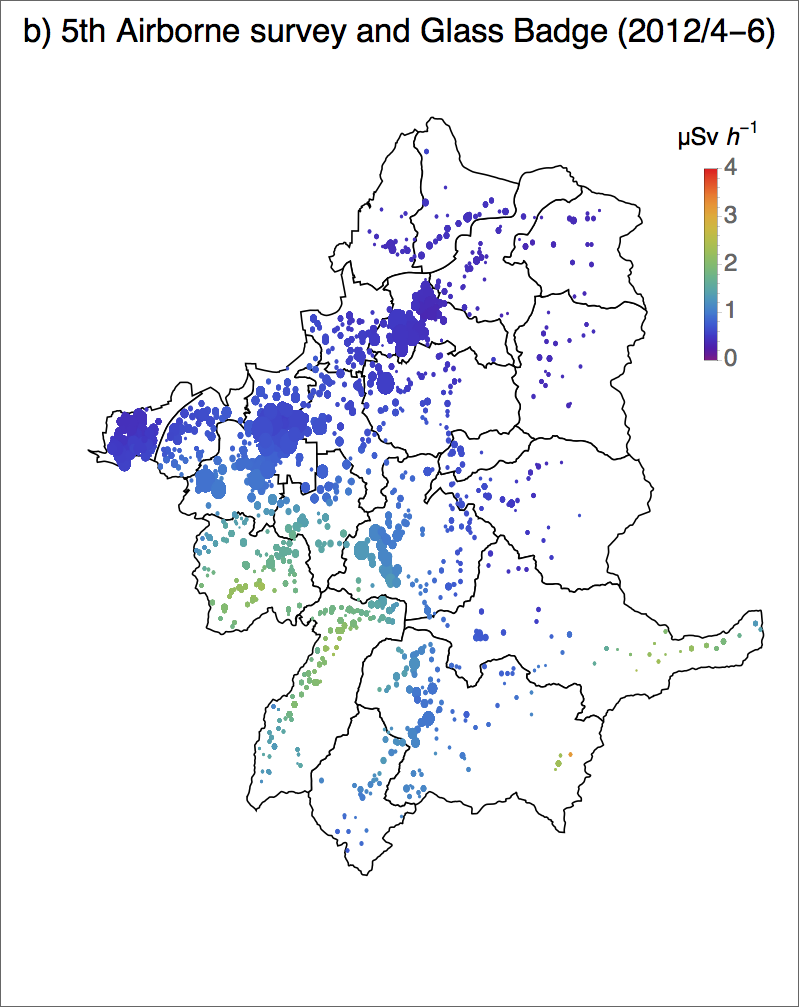}
\includegraphics[width=.78\textwidth]{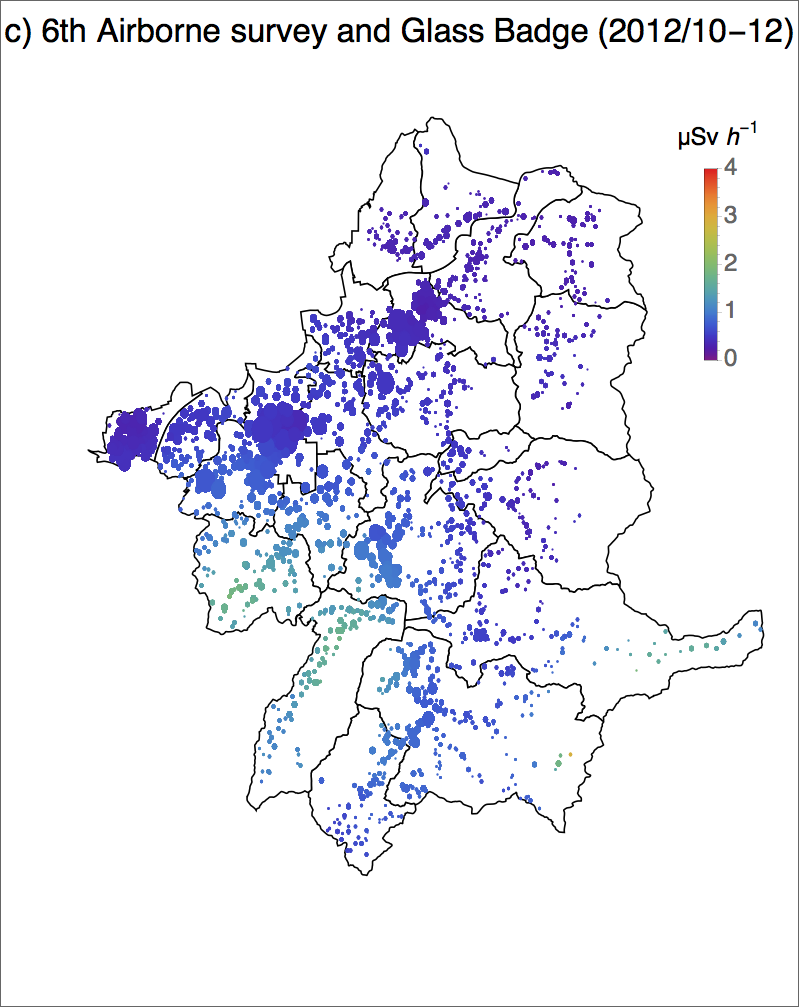}
\end{minipage}
\begin{minipage}{0.44\textwidth}
\includegraphics[width=.78\textwidth]{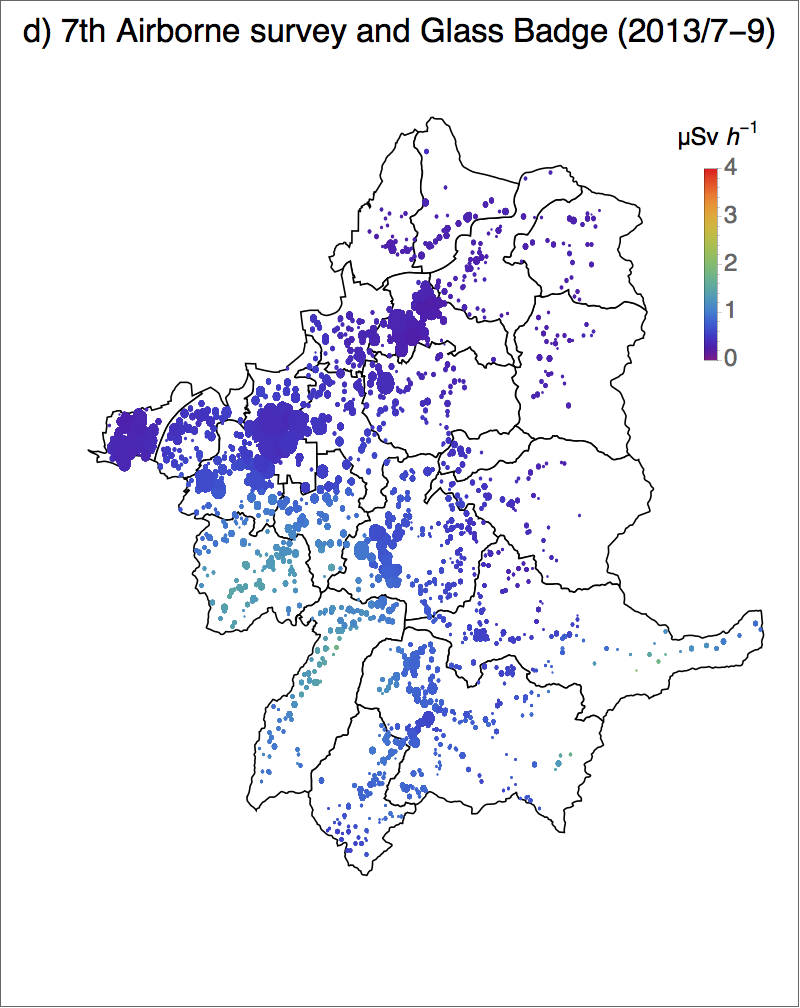}
\includegraphics[width=.78\textwidth]{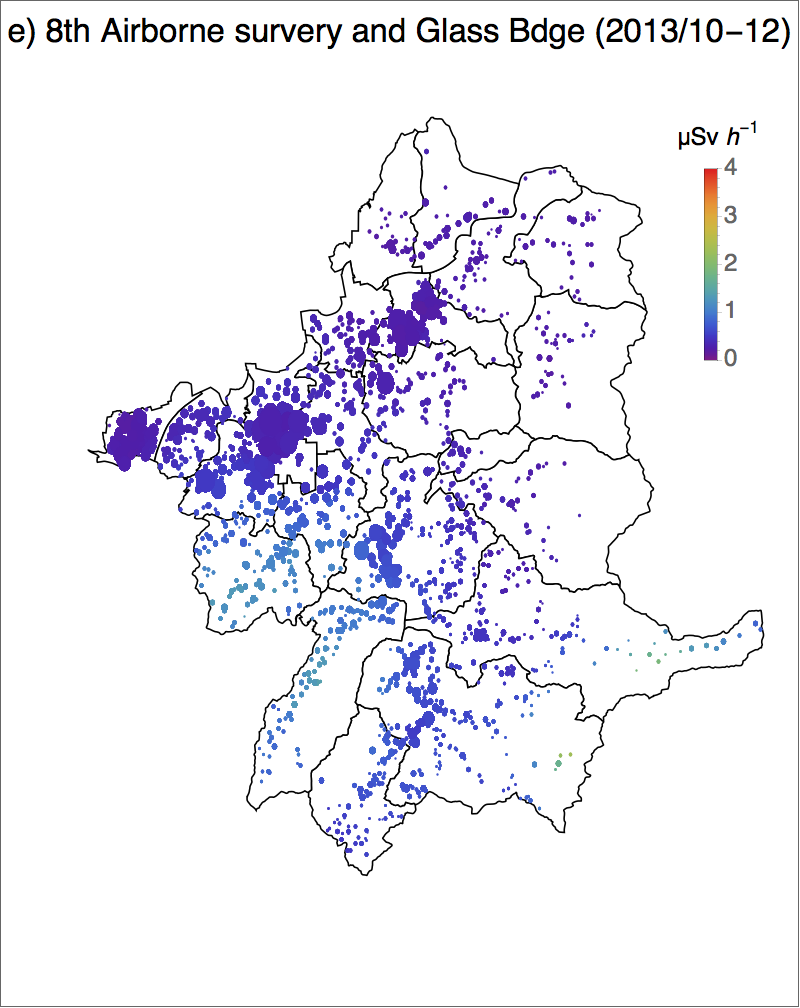}
\includegraphics[width=.78\textwidth]{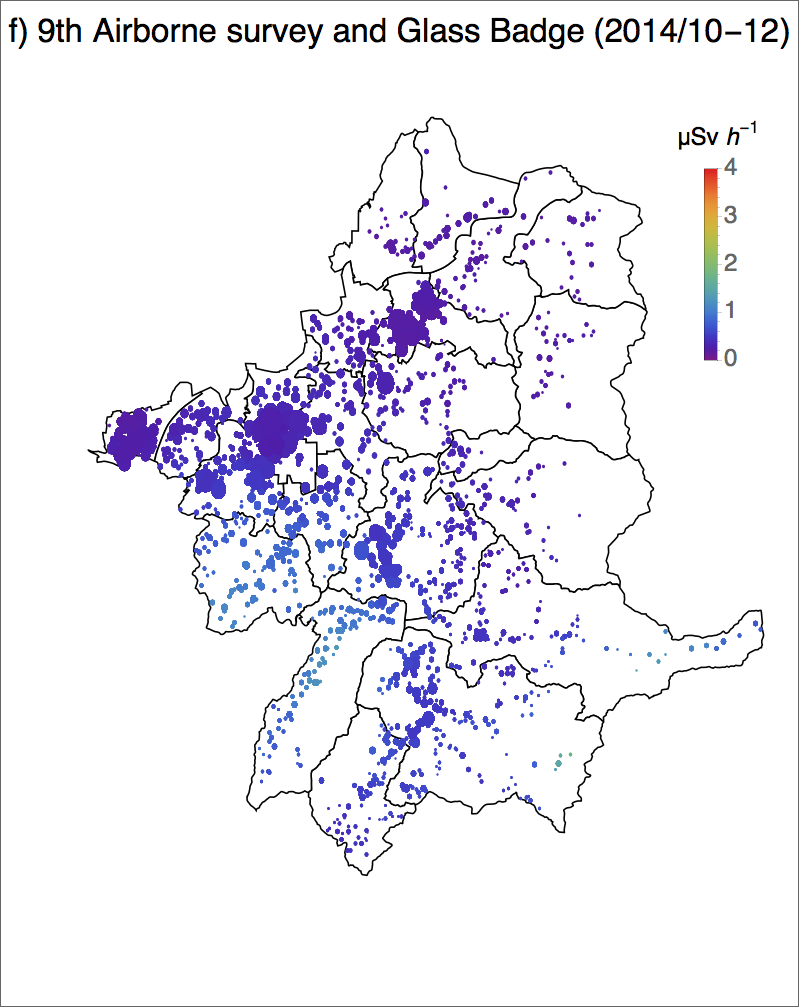}
\end{minipage}
\end{center}
\caption{\label{fig3} Geographical locations of the survey participants and their grid dose rates for each of the six measurement periods. The left-hand column a)-c) shows the 4th, 5th and 6th airborne monitoring, the right-hand column d)-f) shows the 7th, 8th and 9th airborne monitoring. The grid dose rates are color coded ($0-4\, \mu$Sv\,h$^{-1}$), and the area of each circle is drawn proportionally to the number of participants within the grid. In the period corresponding to the 6th airborne monitoring, glass badges were distributed to all citizens, so that c) indicates the overall population distribution within the city.}
\end{figure}

The Japanese government has been continuously monitoring the accident-affected areas using aircraft equipped with radiation monitors flown at a typical altitude of 300~m ~\cite{sanada2014}.
The counting rates are converted to  $\dot{\rm H}^*(10)$  at 1~m above the ground, and the results are published as  average values within a $250\, \rm m \times 250\, m$ grid. Tables listing  $\dot{\rm H}^*(10)$  at each longitude-latitude coordinate of the grid have been disseminated and are available on the internet~\cite{airborne}.

The airborne survey data were then compared with the glass-badge-derived individual doses. As shown in Table 1,  six airborne surveys (4th through 9th)  match the glass-badge measurement periods. We used these six pairs of data in our analysis.

For each period and for each participant, we used the GIS information, the longitude-latitude coordinates of the residence of the participant provided by the Date City office, to look up the $\dot{\rm H}^*(10)$ of the nearest grid point, which we hereafter denote the {\em grid dose rate}, in the corresponding airborne survey database.

In Figure 3, we present the geographical locations of the participants and their grid dose rates for each of the six measurement periods, in separate panels. The grid dose rates are color coded, and the area of each circle is drawn proportionally to the number of participants within each the grid cell.

Since {\em background radiation has been subtracted} from the glass badge data, we subtracted 0.04\,$\mu$Sv\,h$^{-1}$ from the airborne $\dot{\rm H}^*(10)$ data in our analyses to compensate. This value (0.04) was used by the Japanese government in the published formula which related  the ambient dose rate to the effective dose~\cite{safetyassessment}.

\section{RESULTS}

\begin{figure}
\begin{center}
\begin{minipage}{0.49\textwidth}
\includegraphics[width=\textwidth]{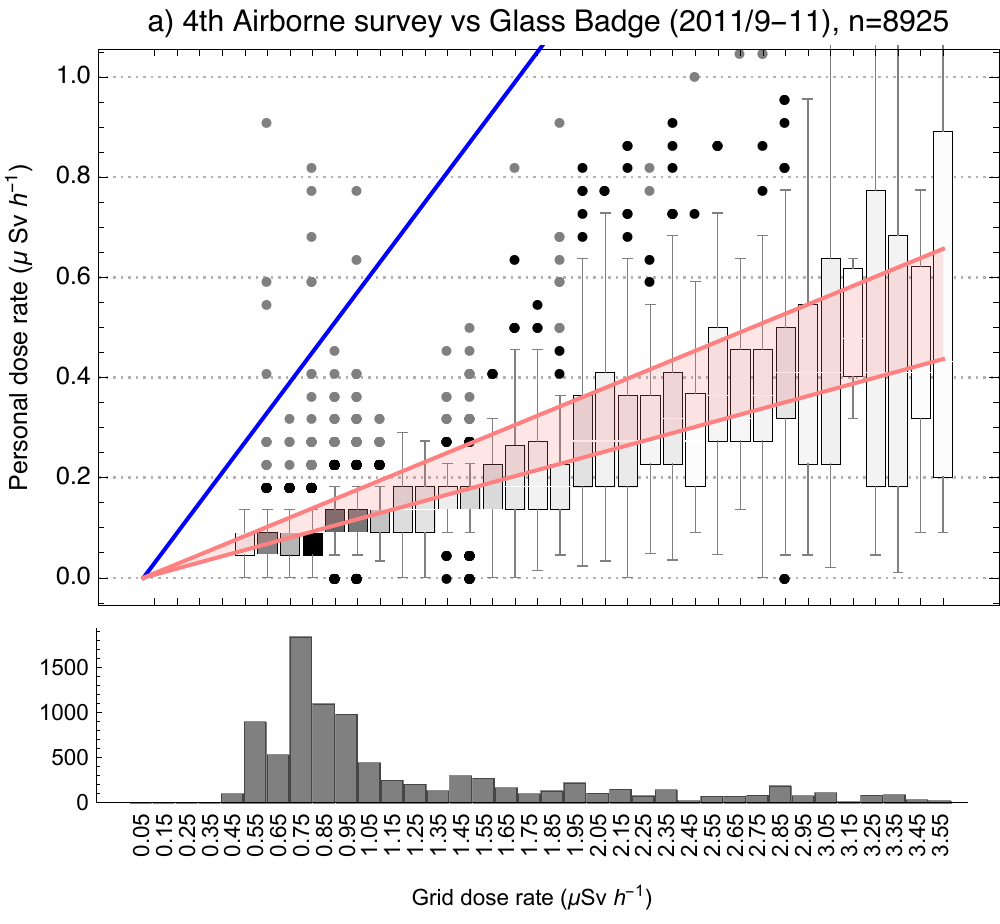}
\includegraphics[width=\textwidth]{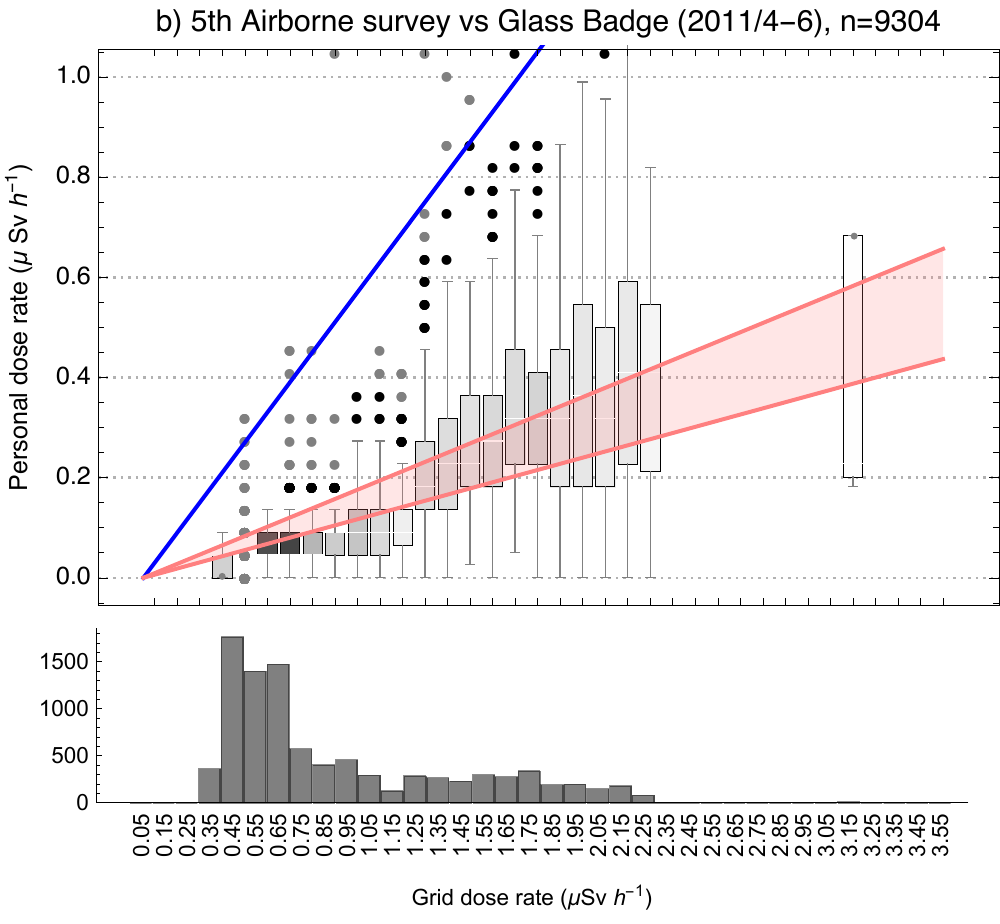}
\includegraphics[width=\textwidth]{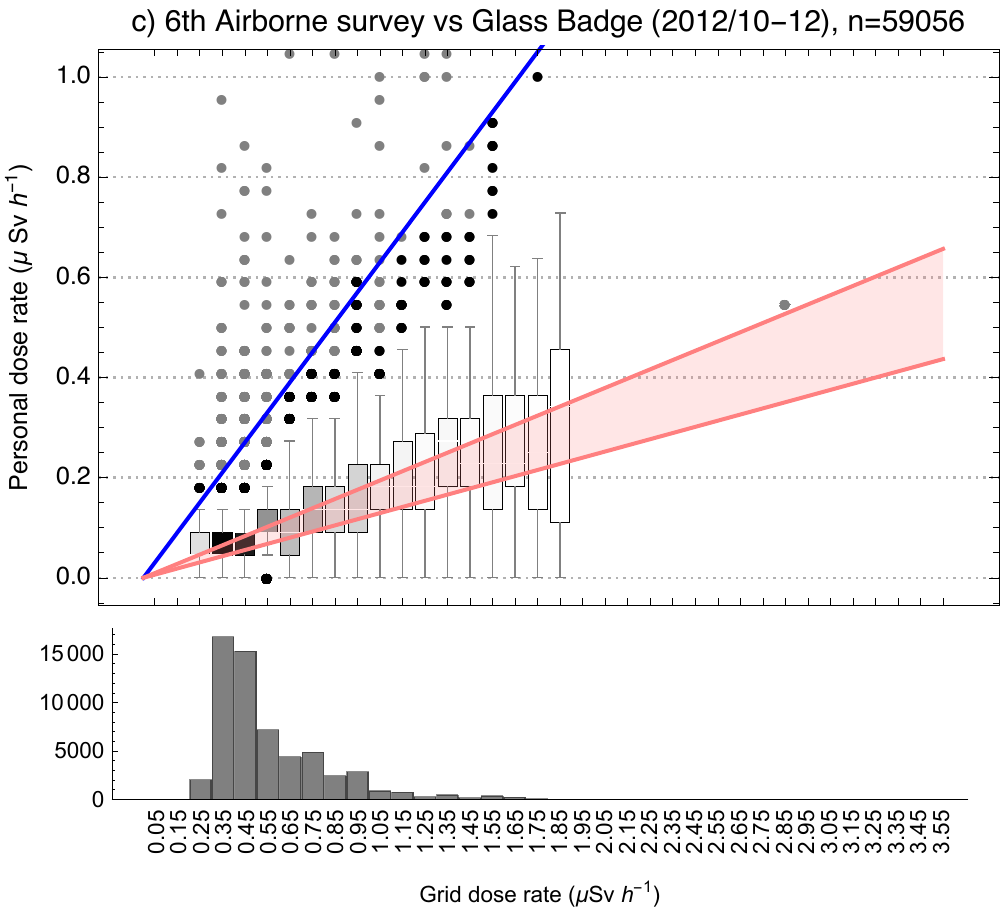}
\end{minipage}
\begin{minipage}{0.49\textwidth}
\includegraphics[width=\textwidth]{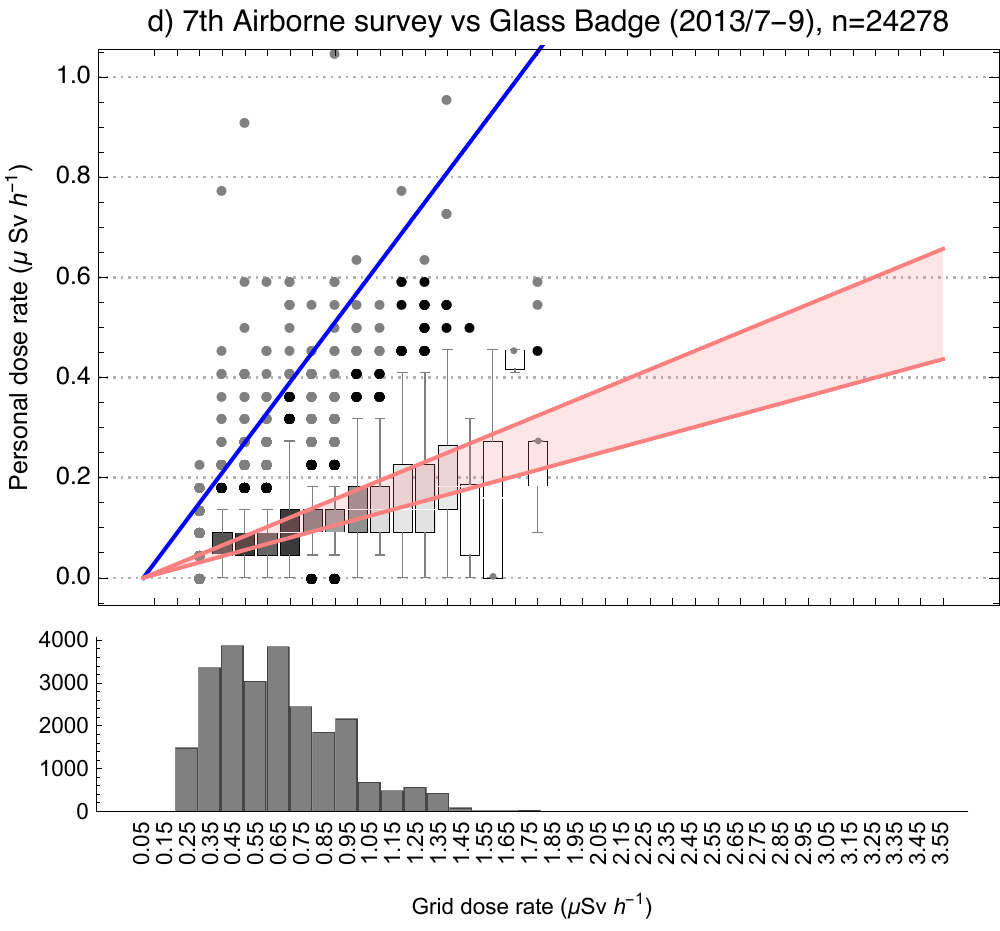}
\includegraphics[width=\textwidth]{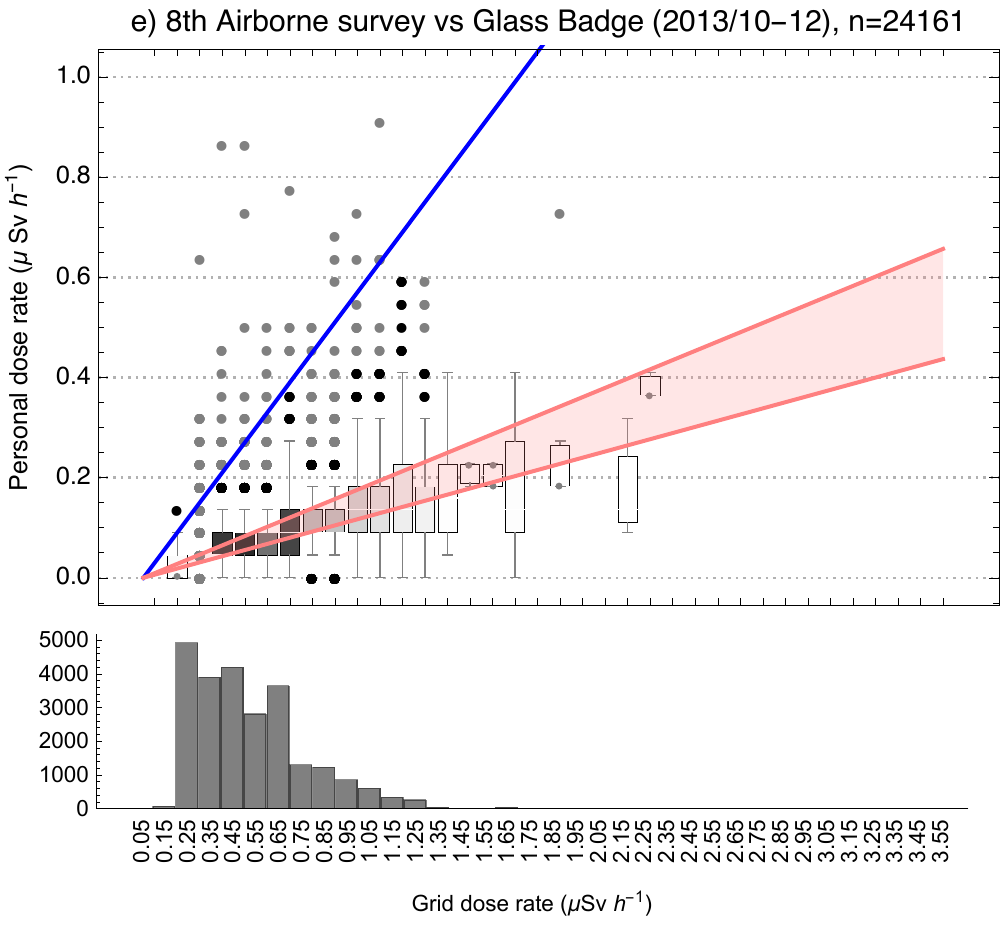}
\includegraphics[width=\textwidth]{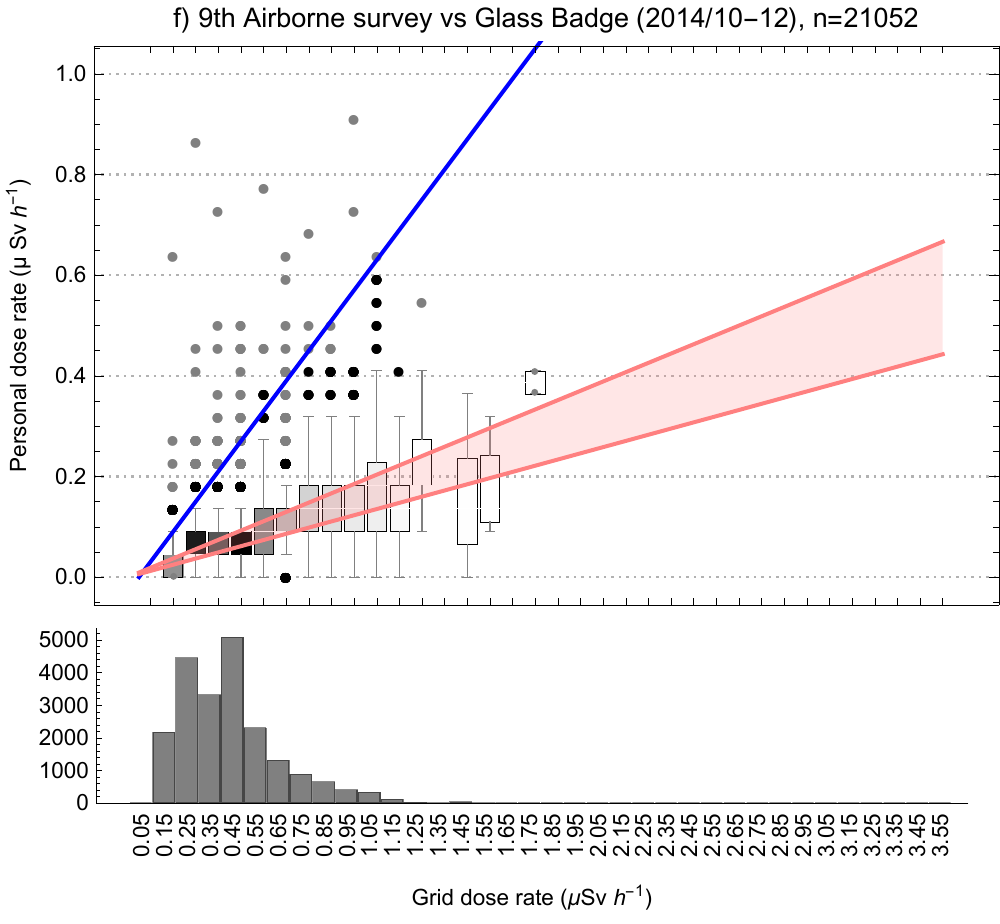}
\end{minipage}
\end{center}
\caption{\label{fig4} 
Box-and-whisker plots of the individual dose rates versus grid dose rates  for the six  periods. The left-hand column a)-c) shows the data corresponding to the 4th, 5th and 6th airborne survey periods, and the right-hand column d)-f) shows  those corresponding to the 7th, 8th and 9th periods.
}
\end{figure}

Figure \ref{fig4} shows the box-and-whisker plots of the individual dose rates versus grid dose rates  for six  periods. The left-hand column from top shows the 4th, 5th, and 6th airborne monitoring, the right hand column from top shows the 7th, 8th, and 9th airborne monitoring (also see Table 1). 
The abscissa is the grid dose rate in increments of $0.1\, \mu$Sv\,h$^{-1}$. The box-and-whiskers represent distributions of the individual dose rates (derived from the three-month accumulated doses) of the subjects, whose grid dose rates were within the bin.  The boxes cover 25th percentile to 75th percentile of the distribution, and the whiskers cover the 1st percentile to 99th percentile of the distribution. The dots represent outliers. Below each box-and-whisker plot, a histogram of population  versus grid dose rate is shown.

The reduction of grid dose rates from the 4th monitoring to the 9th monitoring is already evident in  Fig.~\ref{fig3}.  Correspondingly, the distributions of the number of subjects versus gird dose rates shown in Fig.~\ref{fig4} is steadily shifted to the left (i.e., to lower dose rates).

The mean coefficient $\left<c\right>$, the average ratio of the individual  dose rate to the grid dose rate was
\begin{eqnarray}
\left<c\right>\equiv \left<\frac{\mbox{individual  dose rate}}{\mbox{grid dose rate}}\right> = 0.15\pm 0.03, \label{eq:correlation}
\end{eqnarray}
where $\left < \right>$ denotes the average over all data in the six periods excluding the outliers.
This coefficient was used to draw the pink shaded band in each panel of Fig.~\ref{fig4}. 
Note that the Japanese government employs the following standard behavioral scenario in  evaluating the additional effective dose based on the ambient dose rate~\cite{safetyassessment}:  A person spends 8 hours outdoors without any shielding, and stays 16 hours in a wooden house where the dose rate is 40\% of that of outside
~\cite{tecdoc1162}, with the  natural background dose rate of $0.04\,\mu \rm Sv\,h^{-1}$.
In this scenario, the conversion factor for obtaining the additional effective dose from ambient dose rate excluding the background dose rate is 0.6,  i.e.,
\begin{equation}
\mbox{individual dose rate~}(\mu \mbox{Sv h}^{-1}) = 0.04+ 0.6 \times \mbox{ambient dose rate}.\label{eq:jgf}
\end{equation}
The thick blue line in each panel of Fig.~\ref{fig4} corresponds to this scenario. As shown, the blue line (the government estimate based on its standard behavioral scenario) lies well above the pink band, which is based on  actual measurements.

\begin{figure}
\begin{center}
\includegraphics[width=0.5\textwidth]{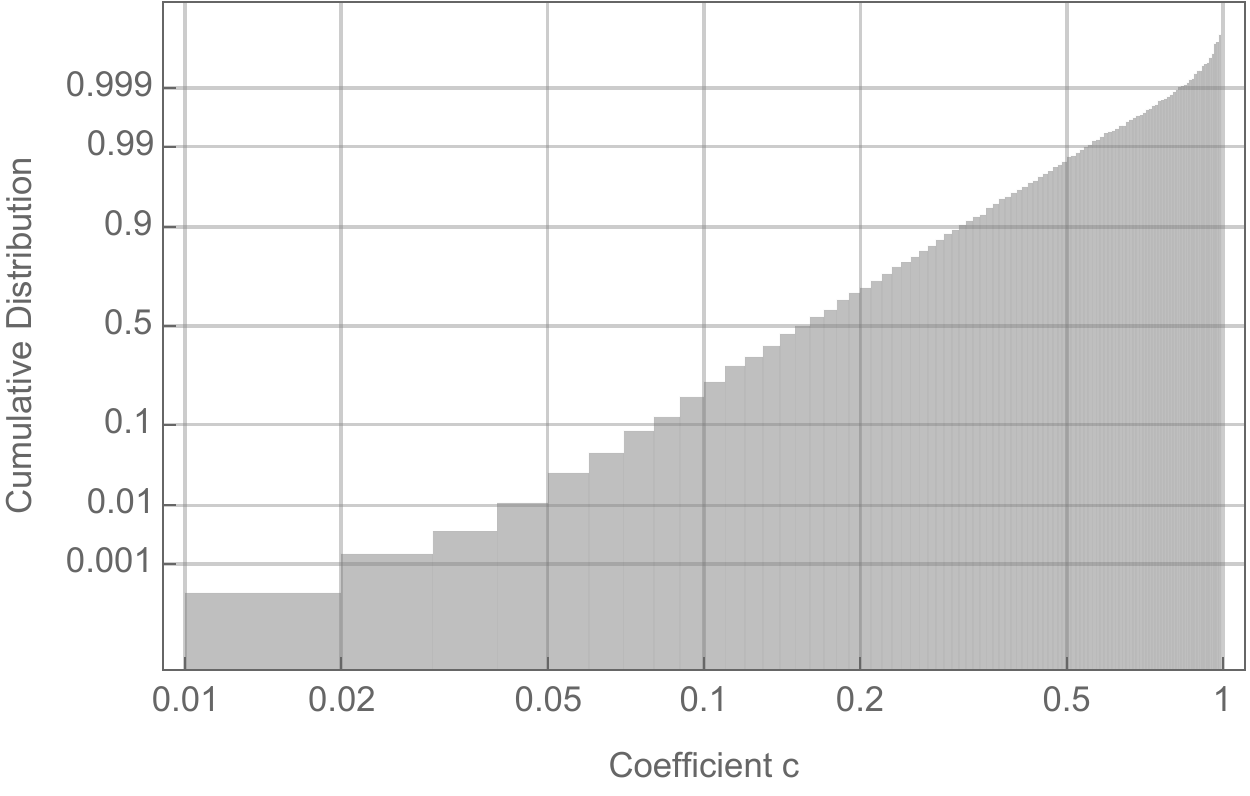}
\end{center}
\caption{\label{fig5} A log-normal plot of the cumulative probability distribution of the $c$ coefficient.}
\end{figure}

The coefficient $c$ has a distribution as shown in Fig.~\ref{fig5}, in which a cumulative probability distribution of $c$ is plotted using a log-normal grid. As shown, 50-percentile is $c=0.15$, 90-percentile is $c=0.31$ and  99-percentile is $c=0.56$. The percentage of the participants whose $c$ coefficient exceeded the Japanese-government value of 0.6 was 0.7\%.

\section{DISCUSSION}

Date City has been monitoring the individual doses of its citizens since five months after the Fukushima-Daiichi Nuclear Power Plant Accident.  As a result, the accumulated individual monitoring data compiled by Date City is the most comprehensive record of the temporal and locational changes in individual doses of residents living in areas affected by the accident. This detailed archive of dose information meets the recommendations of the ICRP: ``In a situation of long-term contamination, it is essential to establish a radiation monitoring system allowing follow-up of the radiological situation and the implementation of adequate protection strategies~\cite{icrp111}.'' The authors analyzed the data upon the request of Date City so that crucial lessons applicable to the post-accident situation can be derived, particularly regarding existing exposure situations, and to help understand what will be needed in the event of future radiological disasters.

The present study showed that the individual dose rate is proportional to the ambient dose rate with a coefficient $c=0.15\pm 0.03$. 
The coefficient obtained in this study is smaller than the government-adopted value of 0.6 by a factor of $\sim 0.25$. Naito et al.~\cite{naitoplosone}, who  targeted a small group of some 100 people,  found a coefficient of $\sim 0.2$, also smaller than the government value.
These results show that  coarse-grained airborne data can be a useful estimator for predicting the individual doses of residents living in contaminated areas.

In our (and Naito et al.'s) work, we both used the actually measured H$^*(10)$ (airborne monitoring) and Hp(10) (individual dose monitoring) data to empirically deduce the coefficient $c$. On the other hand, the Japanese government in 2011 constructed a model to estimate the effective dose from H$^*(10)$, which,  in retrospect, overestimates the personal dose by a factor of $3\sim 5$.

After  large-scale radioactive contamination events such as the FDNPP accident, it should be expected that the response capacity of  local governments will be overstretched. In the case of Fukushima, local governments concentrated their efforts on preparing and distributing the equipment to residents, and it was not possible to ensure the proper use of the personal dosimeters by everyone, as is regularly done for radiation workers. As a result, the  data collected by Date City does not include information about whether or not each participant actually lived at the address registered at the city office, or whether they routinely wore the glass badge as requested. Therefore, strictly speaking, the values measured by the glass-badges are not necessarily equal to the individual doses of every participant. This is a limitation of the present study. However, we believe the differences in actual dosimeter use patterns among the participants do not greatly affect the present results, as discussed below.

Nomura et al.~\cite{nomurabmj2015} analyzed the results of the glass badge measurements conducted continuously for school children in Minamisoma City, and reported that there is no statistically significant difference in the personal doses between the group who reported in the questionnaire that they wore the badge properly, versus the group who said they did not.

In the study of Naito et al.~\cite{naitoplosone}, participants were given strict instructions regarding the usage of the dosimeter, which were duly observed, and they each carried a GPS receiver together with the electronic dosimeter which recorded hourly doses. The hourly dose was compared with the ambient dose rate estimated by combining the GPS information and the airborne monitoring database.  The coefficient deduced from this well-controlled academic research targeting a small number of participants is extremely close to that derived in the current study, based on a municipality-driven long-term monitoring program involving tens of thousands of residents.

\section{CONCLUSIONS}

We have shown that the personal doses measured using glass badges in Date city are proportional to the ambient dose rates at the residences of the participants estimated from  aircraft monitoring. As a result, the authors conclude that it is possible to predict the external exposure dose received by each individual based on the aircraft monitoring data. The conversion factor derived in the present study, should help greatly in estimating the individual external exposure doses in real life in the contaminated areas affected by the FDNPP accident. 
The method obtained in this study could aid in the prediction or in the estimation of
the external doses of residents in the early phase of future radiation accidents involving large-scale contamination.

\begin{acknowledgments}

The authors are grateful to Chiyoda Technol Corporation, and Dr. J. Tada of Radiation Safety Forum for valuable discussions.

\end{acknowledgments}


\begin{thebibliography}{99}
\bibitem{icrp103} ICRP Publication 103. The 2007 Recommendations of the international commission on radiological protection. Ann. ICRP 37 (2-4), 2007.

\bibitem{icrp111} ICRP Publication 111. Application of the commission's recommendations to the protection of people living in long-term contaminated areas after a nuclear accident or a radiation emergency. Ann. ICRP 39 (3), 2009.

\bibitem{ishikawa}  Ishikawa T, Yasumura S, Ozasa K, Kobashi G, Yasuda H, Miyazaki M, Akahane K, Yonai S, Ohtsuru A and Sakai A 2015 The Fukushima Health Management Survey: estimation of external doses to residents in Fukushima Prefecture Scientific reports 5 {\bf 5}:12712

\bibitem{fukushimagb} 2015 Result of the Fukushima City glass badge survey (Japanese).  \url{http://www.city.fukushima.fukushima.jp/uploaded/attachment/50133.pdf}, last accessed July 1, 2016.

\bibitem{tsubokura2015} Tsubokura M, Kato S, Morita T, Nomura S and Kami M 2015 Assessment of the annual additional effective doses amongst Minamisoma children during the second year after the Fukushima Daiichi Nuclear Power Plant Disaster, {\it PloS One} {\bf 10}: e0129114.



\bibitem{evacuation} Ministry of Economy, Trade and Industry web site on evacuation areas \url{http://www.meti.go.jp/english/earthquake/nuclear/roadmap/evacuation_areas.html}.

\bibitem{datecityweb} Date City Report since 2011.3.11 \url{http://www.city.date.fukushima.jp/soshiki/9/7146.html} (Japanese)

\bibitem{datefukko} Date city Fukko-Saisei News Vol 8.  \url{http://www.city.date.fukushima.jp/uploaded/attachment/10035.pdf} (in Japanese)

\bibitem{iaea} International Atomic Energy Agency. Radioactivity in the environment. In: IAEA The Fukushima Daiichi Accident. Technical Volume 4; Radiological Consequences. International Atomic Energy Agency, Vienna International Centre, Vienna, Austria, 2015; 103--105.



\bibitem{nramonitoring} 
Nuclear Regulation Authority, Japan. Airborne Monitoring Survey Results.  \url{http://radioactivity.nsr.go.jp/en/list/307/list-1.html}.

\bibitem{nraairborne} Extension site of Distribution Map of Radiation Dose, etc., \url{http://ramap.jmc.or.jp/map/eng/}.


\bibitem{datestat} Date city statistics (in Japanese) \url{http://www.city.date.fukushima.jp/soshiki/3/4313.html}

\bibitem{datedecontam1} Date City Decontamination Plan v1 (Japanese) \url{http://www.city.date.fukushima.jp/uploaded/attachment/1964.pdf} (Japanese)


\bibitem{datedecontam2} Date City Decontamination Plan v2 (Japanese)  \url{http://www.city.date.fukushima.jp/uploaded/attachment/1963.pdf} (Japanese)

\bibitem{safetyassessment} Safety assessment study committee and environmental recovery review meeting Joint Study Group (Ministry of the Environment meeting, Oct 10, 2011, in Japanese) \url{http://josen.env.go.jp/material/session/joint_001.html}

\bibitem{icru74} ICRU Report 74 (2005) on Patient Dosimetry of X Rays used in Medical Imaging, {\it Journal of the ICRU} {\bf 5}: i

\bibitem{hirayama2013} Hirayama H 2013 An evaluation of personal dosimeter for widely distributed 134 Cs and 137 Cs by using EGS code {\it Radioisotopes} {\bf 62} 335--345 (in Japanese)

\bibitem{nomura2015} Nomura S, Tsubokura M, Hayano R, Furutani T, Yoneoka D, Kami M, Kanazawa Y and Oikawa T 2015 Comparison between Direct Measurements and Modeled Estimates of External Radiation Exposure among School Children 18 to 30 Months after the Fukushima Nuclear Accident in Japan {\it Environ Sci Technol} {\bf 49} 1009--1016

\bibitem{sanada2014} Sanada Y, Sugita T, Nishizawa Y, Kondo A and Torii T 2014 The aerial radiation monitoring in Japan after the Fukushima Daiichi nuclear power plant accident {\it Progress in Nuclear Science and Technology} {\bf 4} 76–80

\bibitem{airborne} Nuclear Regulation Authority. Airborne Monitoring Survey Results (including CSV files). \url{http://radioactivity.nsr.go.jp/ja/list/362/list-1.html} (Japanese)


\bibitem{tecdoc1162} IAEA: Generic procedures for assessment and response during a radiological emergency, TECDOC-1162, Table E4 (2000)

\bibitem{naitoplosone} Naito W, Uesaka M, Yamada C, Kurosawa T, Yasutaka T, Ishii H 2016 Relationship between Individual External Doses, Ambient Dose Rates and Individuals' Activity-Patterns in Affected Areas in Fukushima following the Fukushima Daiichi Nuclear Power Plant Accident, {\it PLoS ONE} {\bf 11}: e0158879.

\bibitem{nomurabmj2015} Nomura S, Tsubokura M, Hayano R and Yoneoka D 2015 Compliance with the proper use of an
 individual radiation dosimeter among
 children and the effects of improper use
 on the measured dose: a retrospective
 study 18–20 months following Japan's
 2011 Fukushima nuclear incident, {\it BMJ} open {\bf 5}:e009555.


\end{thebibliography}
\end{document}